\documentclass[twocolumn,prl,aps,showpacs]{revtex4}
\usepackage{times}
\usepackage{bm}
\usepackage{epsfig}

\begin{document}

\preprint{PRL/electron-phonon}

\title{Direct temporal measurement of hot-electron relaxation in a phonon-cooled metal island}

\author{D. R. Schmidt}
\author{C. S. Yung}
\author{A. N. Cleland}
\email[]{cleland@physics.ucsb.edu}
\affiliation{Department of Physics and iQUEST, University of
California at Santa Barbara, Santa Barbara, CA 93106}

\date{\today}

\begin{abstract}

We report temporal measurements of the electronic temperature and the electron-phonon thermal
relaxation rate in a micron-scale metal island, with a heat capacity 
of order 1 fJ/K ($C
\sim~$10$^{7}$ $k_{B}$) .  We employed a superconductor-insulator-normal metal tunnel junction,
embedded in a radio-frequency resonator, as a fast ($\sim$ 20 MHz) thermometer.  A resistive heater
coupled to the island allowed us to drive the electronic temperature well above the phonon
temperature.  Using this device, we have directly measured the thermal relaxation of the hot
electron population, with a measured rate consistent with the theory for dynamic electron-phonon
cooling.
\end{abstract}


\maketitle

\begin{figure}
\centerline{\epsfig{file=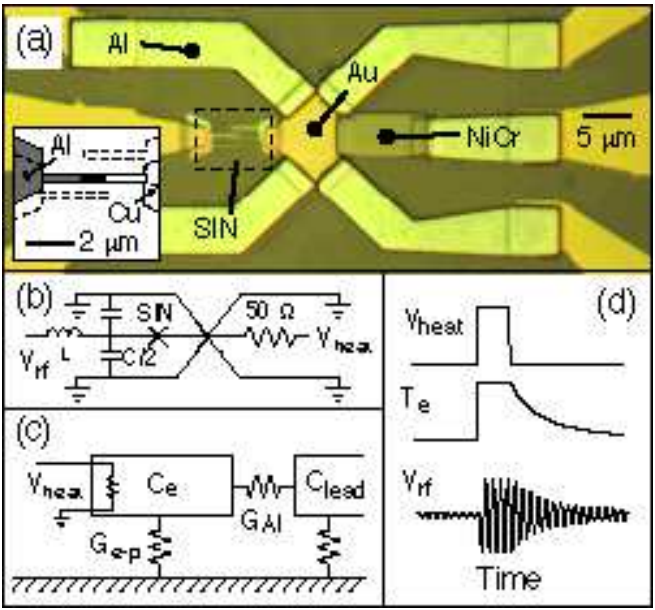, width=1.0\linewidth}} \caption{
 (Color Online)(a) Optical micrograph of the electron calorimeter.  The center Au
 island is contacted on the left by a rf-SIN thermometer,
 and on the right by a NiCr resistor.  The outer ground leads and the
 contact right of the resistor are superconducting Al. {\em inset}:
 Detail drawing of the SIN junction, Al shown in gray, Cu in white, and
 overlap junction area in black. The dotted outlines are fabrication artifacts.
 (b) Electrical circuit.  The SIN thermometer is embedded in an $LC$
 resonator formed by a discrete inductor and the stray lead
 capacitance.  The junction resistance is monitored
 using the power reflected from the $LC$ resonator at its resonance frequency.
 (c) Thermal schematic.  The calorimeter electron gas $C_{e}$ is
 thermally isolated by the superconducting Al
 contacts ($G_{Al}$); the dominant thermal link is thus through the substrate
 phonons ($G_{e-p}$). The NiCr resistor
directly heats the electron gas. (d) Timing diagram. The voltage pulse applied to the heater causes
the temperature to rise, saturate, and then decay. The envelope of the reflected power from the
$LC$ resonator is directly related to the temperature. } \label{fig:device}
\end{figure}

Measurement of the heat capacity $C$ of a thermodynamic system, in contact with a thermal
reservoir through a thermal conductance $G$, necessitates the measurement of temperature over time
scales shorter than the characteristic thermal relaxation time $\tau = C/G$. For mesoscopic
devices, this time scale becomes exceedingly short, as both the electron and phonon heat capacities
scale with device volume $V$. Furthermore, it is difficult to thermally isolate a phonon system
from its environment, as even a very weak mechanical suspension is limited at low temperatures by
the scale-independent quantum of phonon thermal conductance
\cite{Rego:1998,Schwab:2000,Yung:2002,Cleland:2001}. Electrons in a metal however naturally
decouple from their phonon environment at low temperatures, with an electron-phonon thermal
conductance $G_{e-p} \propto V T^{4}$. As the electron heat capacity scales as $C_{e} \propto V T$,
the electron-phonon thermal relaxation time $\tau_{e-p} = C_{e}/G_{e-p}$ is independent of volume,
and scales as $T^{-3}$. At 1 K, $\tau_{e-p}$ is of order 10 nanoseconds, a time scale that is
accessible using a radio-frequency superconductor-insulator-normal metal (rf-SIN) tunnel junction
thermometer \cite{Note:Schmidt:2003}.

In this letter, we present large-bandwidth measurements of the electronic temperature of a
micron-scale metal island.  Our measurement has ample bandwidth with which to directly measure
$\tau_{e-p}$ at temperatures up to 1 K. This system therefore allows us to probe the thermodynamic
behavior of electrons in very small metal volumes, potentially with heat capacities as small as
$10~k_B$. Such small metal volumes are prime candidates for energy absorbers in far-infrared
photon-counting bolometers \cite{Yung:2002}, and would allow unprecedented calorimetric sensitivity
in the mesoscopic regime. Measurements over time scales shorter than $\tau_{e-p}$ are also critical
for developing a complete understanding of the thermodynamics of mesoscopic systems.

The thermal decoupling of electrons and phonons at low temperatures was first described
theoretically by Little \cite{Little:1959}, with a more general discussion provided by Gantmakher
\cite{Gantmakher:1974}. For a bulk metal with volume $V$, the power flow $P_{e-p}$ from the
electron gas at temperature $T_e$ to the phonon gas at $T_p$ is given by
\begin{equation}
    P_{e-p} = \Sigma V ( T^{n}_{e}-T^{n}_{p}),
\label{eq:Pep}
\end{equation}
where $\Sigma$ is a material-dependent parameter. For a spherical Fermi surface and a Debye phonon
gas, $n = 5$.

A number of researchers have verified that Eq. (\ref{eq:Pep}) applies to the static heating of
thin-film metals, albeit with $n$ slightly lower than 5 (fit values for $n$ fall in the range from
4.5 to 4.9, with values for $\Sigma$ in the range $1-2 \times 10^{9}$ W/m$^3$-K$^5$
\cite{Roukes:1985,Wellstood:1994}). These measurements were made using a dc superconducting quantum
interference device (dc-SQUID) to measure the Johnson-Nyquist noise in the metal film, and thus
extract the electronic temperature.

A second approach to measuring the electron temperature in thin metal films was presented by Nahum
\emph{et al.} \cite{Nahum:1993}: using a SIN tunnel junction as an electronic thermometer. These
authors suggested that such a structure could form the heart of a bolometric detector. Measurements
of the static energy distribution of electrons in a normal metal under voltage bias were made by
Pothier \emph{et al.} using a similar SIN-based thermometer \cite{Pothier:1997}. Yung \emph{et al.}
\cite{Yung:2002} also demonstrated a SIN thermometer in contact with a normal metal island, the
whole fabricated on a micron-scale suspended GaAs substrate.

Here we study the \emph{dynamic} temperature response of a small metal island using a SIN tunnel
junction thermometer. Well below the superconducting transition temperature $T_{C}$, the tunnel
junction's small-signal resistance at zero bias, $R_0 \equiv dV/dI(0)$, is exponentially dependent
on the ratio of temperature $T$ to the superconducting energy gap $\Delta$, $R_0 \propto
e^{\Delta/k_{B}T}$.  A sub-micron scale SIN tunnel junction therefore has a low-temperature
resistance that can easily exceed $10^6~\Omega$, limiting conventional time-domain measurements to
bandwidths of order 1 kHz. In order to monitor changes in this resistance at sub-microsecond time
scales, we circumvent the unavoidable stray capacitance in the measurement circuit by embedding the
junction in a $LC$ resonant circuit, as shown in Fig. \ref{fig:device} \cite{Note:Bandwidth}. We
then measure the resistance of the SIN junction, and thus the normal metal electron temperature, by
measuring the power reflected from the circuit at the $LC$ resonance frequency. A change of the
junction resistance $R_{0}$, induced by heating the electrons, in turn changes the amplitude of the
reflected radio frequency signal. In this technique, the stray cable capacitance $C$ is in
resonance with the inductor $L$, the resonator also serving to impedance-match the resistance of
the tunnel junction to the measurement system.  This readout scheme is analogous to that employed
in the radio-frequency single electron transistor (rf-SET) \cite{Schoelkopf:1998}.

Our device is fabricated on a 4$\times$4$\times$0.5 mm$^3$ single-crystal GaAs chip using four
lithography steps. A 85 nm thick Au center island and wire-bond pads were first deposited on the
GaAs substrate; an intermediate Au pad was also deposited in this layer. We then deposited a 100 nm
thick NiCr film, designed to have a 50 $\Omega$ resistance, matching the characteristic cable
impedance $Z_0$. The ground leads and heater contact were evaporated in the third layer, using
superconducting Al to ensure thermal isolation below 1 K \cite{note:AlG}. The NiCr contacts the Al
\emph{via} the intermediate Au pad, to ensure low interfacial resistivity. The tunnel junction
thermometer was deposited in the fourth lithography step.  We used a standard suspended resist
bridge, double-angle evaporation method to define the tunnel junction \cite{Fulton:1987}: A 90 nm
thick Al electrode was evaporated, and the Al then oxidized in 200 mTorr of pure O$_{2}$ for 90 s.
The junction was completed by evaporating a 90 nm thick Cu counterelectrode, which also contacted
the center Au island.

\begin{figure}
\centerline{\epsfig{file=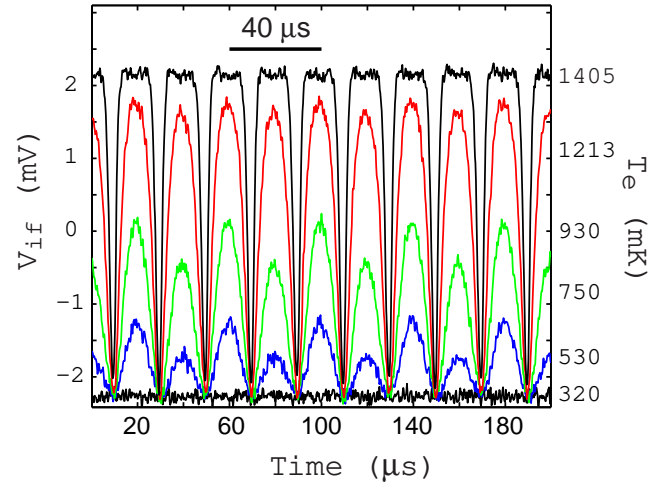, width=1.0\linewidth}} \caption{(Color Online)
Response to a 25 kHz heater drive ({\em left axis}: Mixer if voltage, {\em right axis}: Electronic
temperature).  The thermometer response is at 50 kHz as discussed in the text.  Each trace is the
result of 256 averages with a 2 MHz low-pass filter. The baseline signal is for zero heater power,
with power ranging from 300 pW to 100 nW. At the highest power the signal clips at $T=T_{C}\approx
$ 1400 mK. The 25 kHz components at low power are due to a dc offset in the heater signal. }
\label{fig:lowfreq}
\end{figure}

The device is shown in Fig. \ref{fig:device}(a). Note that the Au center island is electrically
grounded, so that heating signals applied to the NiCr resistor do not couple directly to the SIN
junction, but instead affect it by changing the temperature of the Au island. The heating signals
are in principle therefore limited by diffusion time from the NiCr through the Au island and then
along the Cu electrode to the SIN tunnel junction; we estimate this time to be less than 10 ns.

We mounted the chip containing the device on a printed circuit board, which was enclosed in a brass
box.  Gold  wire bonds (25 $\mu$m diameter) were made between the Au bond pads on the chip and Cu
coplanar striplines on the circuit board.  A chip inductor with $L =$ 390 nH was placed in series
with the SIN junction.  The resonance capacitance $C$ was from the geometric capacitance of the
stripline and Au bond pads, with $C =$ 0.5 pF. The expected $LC$ resonance frequency is $f_{res} =
1/2\pi(LC)^{1/2}\cong$ 350 MHz, the tuned circuit quality factor is $Q = \sqrt{L/CZ_{0}^{2}} \cong
20$, and the measurement bandwidth is $\Delta f = f_{res}/Q \cong$ 20 MHz.  The measurement circuit
is shown in Fig.  \ref{fig:device}(b).  The tunnel junction is configured for simultaneous dc and
rf measurements \emph{via} a bias tee, not shown in Fig.  \ref{fig:device}.

We have described the technical aspects of rf-SIN thermometry elsewhere \cite{Note:Schmidt:2003}.
Here we will describe the salient aspects as they pertain to these measurements.  We determined the
resonance frequency of the $LC$ circuit to be 345 MHz.  A carrier signal source was connected
through a directional coupler to a coaxial line, which was in turn connected to the $LC$ resonant
circuit. The carrier frequency was set close to the $LC$ resonant frequency \cite{Note:Power}.  The
signal reflected from the $LC$ resonator was high-pass filtered and amplified using a low-noise
amplifier.  This amplified signal was then mixed with a local oscillator (lo), provided by a second
rf signal source phase-locked to the carrier source.  The intermediate frequency (if) output from
the mixer was low-pass filtered, amplified, and the resulting time-dependent signal captured by a
sampling oscilloscope.  The NiCr resistor was heated using either a dc or an rf pulsed source: A
pulse sent to the resistor heats the NiCr, the Au island and the Cu electrode, changing the
electron temperature, and therefore changing the amplitude of the carrier signal reflected from the
tuned $LC$ circuit, as shown in Fig. \ref{fig:device}(d).

In order to characterize the response of the system, we first heated the NiCr resistor using a $f_0
=$ 25 kHz sinusoidal drive signal. Figure \ref{fig:lowfreq} shows the response for various drive
powers. The if signal was low-pass filtered ($f < $ 2 MHz), and each curve is the result of
averaging 256 drive periods.  The left axis is the mixer if voltage, and the labels on the right
axis indicate the electron temperature inferred from the change in reflected signal.  The
instantaneous power dissipated in the resistor is proportional to the square of the voltage applied
the heater ($P(t) = V^{2}(t)/R_{NiCr}$); this causes the reflected signal to be modulated at twice
the heater signal, $2 f_0 = $ 50 kHz.  At low powers $P$, contributions at 25 kHz were also
present, due to a small dc offset on the heater voltage $V(t)$, $V(t) = V_{dc} + V_0 \sin 2 \pi f_0
t$.  At the highest powers, the reflected signal is clipped near the Al superconducting transition
temperature: The junction resistance is temperature-independent above $T_{C}$.

\begin{figure}
\centerline{\epsfig{file=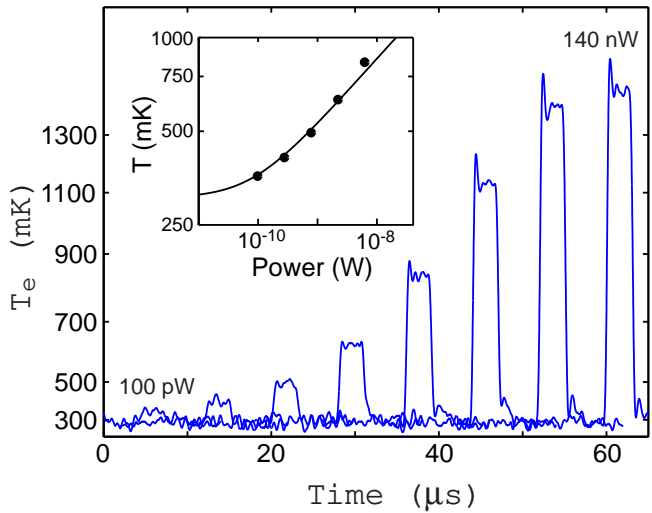, width=1.0\linewidth}} \caption{Composite response to pulsed
heating signals, with pulses 3.0 $\mu$s long with peak power 0.1, 0.3, 0.8, 2.2, 6.2, 17.6, 49.0,
and 140 nW. The resulting electronic temperature for each pulse is used to determine the relation
$P(T_{e},T_{p})$. \emph{Inset:} The solid line is a fit to $P(T_{e},T_{p}) = V\Sigma
(T^{n}_{e}-T^{n}_{p})$, with $T_{p}=$ 300 mK, $n=$ 4.7, $V =$ 10 $\mu$m$^{3}$, and $\Sigma$ =
2.1$\times$10$^{9}$ W/m$^3$-K$^{4.7}$. } \label{fig:pulses}
\end{figure}

The measured signal depends on the proper adjustment of the detection mixer's local oscillator (lo)
phase. In order to correctly adjust this phase, we first applied a heater signal sufficient to get
a clipped response. The phase of the lo was then adjusted to achieve maximum differential response
between the lowest ($\cong$ 300 mK) and highest ($\cong$ 1400 mK) electron temperatures. The SIN
junction ranges from 105 k$\Omega$ to 6 k$\Omega$ over this temperature range, and passes through
the value of $R_{0}$ where optimal matching with the cable impedance occurs \cite{note:ROptimal}.
In the parlance of radio-frequency electronics, the carrier signal is over-modulated, so the
absolute value of the reflected power is a non-monotonic function of temperature. However, as we
are sensitive to the phase of the carrier, the proper quadrature of the mixer if voltage does have
a monotonic response. Finally, the reflected if signal as a function of cryostat temperature, for
no heater voltage applied, was used to construct the temperature calibration, $V_{if}(T)$.

We measured the \emph{quasi-static} relation between the electron-phonon power flow $P_{e-p}$ and
the electron and phonon temperatures, $T_{e}$ and $T_{p}$, as given by Eq.  (\ref{eq:Pep}). We
applied a series of 3 $\mu$s pulses while varying the peak heating power, and monitored the
resulting time-dependent electron temperature.  The substrate temperature was kept at 300 mK. The
signal was filtered with a 2 MHz low-pass filter, and the result of 256 averages is shown in Fig.
\ref{fig:lowfreq}.  This is equivalent to a dc heating measurement with a key difference, namely
that as the heating pulses were delivered to the device at a 1 kHz repetition rate, the duty cycle
was only 0.3\%, so that the substrate phonons did not have sufficient time to heat.  The equivalent
measurement in a dc heating experiment requires 300 times as much power, with significant phonon
heating a likely outcome.  We find a fit relation matching that of Eq. (\ref{eq:Pep}), with $n$ =
4.7 and $\Sigma$ = 2.1$\times$10$^{9}$ W/m$^3$-K$^{4.7}$, in good agreement with previously
measured values \cite{Roukes:1985,Wellstood:1994,Yung:2002}.

We finally performed measurements of \emph{dynamic} electron-phonon cooling, by monitoring the
detailed time-dependent behavior of the electron temperature at the end of a heating pulse. Figure
\ref{fig:pulses} shows the measured response to a heater pulse (2560 averages, using a 50 MHz
low-pass filter).  The heating voltage pulse was configured to have 1.6 ns leading and trailing
edge widths. The initial temperature rise is at least as fast as the time resolution of the
measurement, with an expected rate $\dot{T} = P/C_{e} \cong$ 140 mK/nsec, as we are directly
heating the electron population. The rapid onset also indicates that electron diffusion in the
composite metal structure is not a rate-limiting factor. At the end of the pulse, the heating power
drops to zero, leaving a non-equilibrium hot electron population that relaxes by phonon emission.
Initially this relaxation is seen to be quite rapid, but it slows markedly as the electron
temperature nears the phonon temperature.

\begin{figure}
\centerline{\epsfig{file=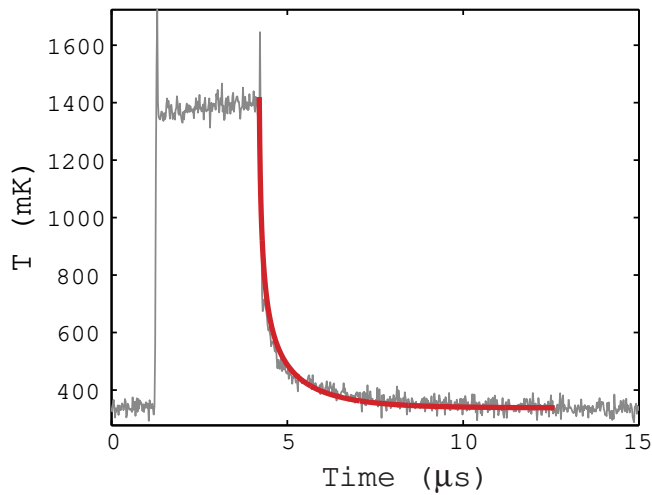, width=1.0\linewidth}} \caption{(Color Online)
Response to a 3.0 $\mu$s pulse with peak power 140 nW applied to the heater. The electronic
temperature quickly rises to $\approx$ 1400 mK at the start of the pulse. The trailing edge of the
response decays with a temperature-dependent relaxation time. The solid line is a fit to the
thermal model discussed in the text. The ringing is due to the if amplifier circuit.}
\label{fig:relax}
\end{figure}

The shape of the relaxation curve shown in Fig. \ref{fig:relax} can be understood by examining the
dynamics of the electron temperature. The electron heat capacity is $C_{e} = \gamma V T_{e}$, where
$\gamma$ is the Sommerfeld constant. The power flow to the phonons is given by Eq. (\ref{eq:Pep}).
The time rate of change of the electron temperature $\dot{T}_e$ is then
\begin{equation}
   \dot{T}_e = -\frac{\Sigma}{\gamma}(T_{e}^{n-1}-\frac{T_{p}^{n}}{T_{e}})
\label{eq:Tdot}.
\end{equation}
Using the normalized temperature $\theta \equiv T_{e}/T_{p}$, this is
\begin{eqnarray}
   \dot{\theta}
   = - \frac{1}{n} \frac{1}{\tau_{e-p}(T_p)} (\theta^{n-1}-1/\theta),
\label{eq:thetadot}
\end{eqnarray}
in terms of the small signal thermal relaxation rate $\tau_{e-p}^{-1} = n \Sigma
T_{p}^{n-2}/\gamma$ for electrons near the phonon temperature \cite{Note:SmallSignal}.  We fit our
measured response to Eq. (\ref{eq:thetadot}) using this rate as the only adjustable parameter,
finding the value $\tau_{e-p} = 1.6~\mu$s \cite{Note:WhichN}. This is in agreement with the
measured value of $\Sigma$, and a composite $\gamma$ which takes into account the relative metal
volumes in the device. We can thus determine the the heat capacity of the 
metal island, $C_e \sim 1$
fJ/K $\cong 10^{7}~k_{B}$ at 300 mK.

There are some extremely interesting opportunities for electronic calorimetry in this temperature
and size regime. Intriguing theoretical results have been presented for the thermodynamic response
of mesoscopic superconducting disks \cite{Deo:2000}, and giant moment electronic paramagnets such
as PdMn \cite{Nieuwenhuys:1975} and PdFe \cite{Peters:1984} offer a means of probing the
thermodynamics of a mesoscopic phonon-electron-spin-coupled system.

We are far from the ultimate calorimetric limits for this technique.  Devices with active metal
volumes that are smaller by a factor of $10^4-10^5$ can be fabricated, yielding a total heat
capacity of order $\sim 10-100~k_{B}$ at 30 mK.  Changes in the heat capacity of less than 10\% are
easily detected, yielding a sensitivity of order 1 $k_B$, i.e. that associated with a single degree
of freedom.

In summary, we have performed sub-$\mu$s timescale measurements of the electron temperature of a
micron scale metal island, cooled dynamically by phonon emission. The ability to apply and measure
the response to fast heat pulses has permitted us to directly measure the electron-phonon thermal
relaxation, and thus extract the heat capacity of the metal island. This, to our knowledge, is the
smallest measured heat capacity to date. The device that we have fabricated is a major step forward
for mesoscopic thermodynamics, provides a platform for sub-aJ/K calorimetry, and can potentially
play an important role in future single photon and phonon bolometers.

\begin{acknowledgments}
We acknowledge financial support provided by the NASA Office of Space Science under grants
NAG5-8669 and NAG5-11426, the Army Research Office under Award DAAD-19-99-1-0226, and the Center
for Nanoscience Innovation for Defense. We thank Bob Hill for processing support.
\end{acknowledgments}


\begin{thebibliography}{22}
\expandafter\ifx\csname natexlab\endcsname\relax\def\natexlab#1{#1}\fi
\expandafter\ifx\csname bibnamefont\endcsname\relax
  \def\bibnamefont#1{#1}\fi
\expandafter\ifx\csname bibfnamefont\endcsname\relax
  \def\bibfnamefont#1{#1}\fi
\expandafter\ifx\csname citenamefont\endcsname\relax
  \def\citenamefont#1{#1}\fi
\expandafter\ifx\csname url\endcsname\relax
  \def\url#1{\texttt{#1}}\fi
\expandafter\ifx\csname urlprefix\endcsname\relax\def\urlprefix{URL }\fi
\providecommand{\bibinfo}[2]{#2}
\providecommand{\eprint}[2][]{\url{#2}}

\bibitem[{\citenamefont{Rego and Kirczenow}(1998)}]{Rego:1998}
\bibinfo{author}{\bibfnamefont{L.~G.~C.} \bibnamefont{Rego}} \bibnamefont{and}
  \bibinfo{author}{\bibfnamefont{G.}~\bibnamefont{Kirczenow}},
  \bibinfo{journal}{Phys. Rev. Lett.} \textbf{\bibinfo{volume}{81}},
  \bibinfo{pages}{232} (\bibinfo{year}{1998}).

\bibitem[{\citenamefont{Schwab et~al.}(2000)\citenamefont{Schwab, Henriksen,
  Worlock, and Roukes}}]{Schwab:2000}
\bibinfo{author}{\bibfnamefont{K.}~\bibnamefont{Schwab}},
  \bibinfo{author}{\bibfnamefont{E.~A.} \bibnamefont{Henriksen}},
  \bibinfo{author}{\bibfnamefont{J.~M.} \bibnamefont{Worlock}},
  \bibnamefont{and} \bibinfo{author}{\bibfnamefont{M.~L.}
  \bibnamefont{Roukes}}, \bibinfo{journal}{Nature}
  \textbf{\bibinfo{volume}{404}}, \bibinfo{pages}{974} (\bibinfo{year}{2000}).

\bibitem[{\citenamefont{Yung et~al.}(2002)\citenamefont{Yung, Schmidt, and
  Cleland}}]{Yung:2002}
\bibinfo{author}{\bibfnamefont{C.~S.} \bibnamefont{Yung}},
  \bibinfo{author}{\bibfnamefont{D.~R.} \bibnamefont{Schmidt}},
  \bibnamefont{and} \bibinfo{author}{\bibfnamefont{A.~N.}
  \bibnamefont{Cleland}}, \bibinfo{journal}{Appl. Phys. Lett.}
  \textbf{\bibinfo{volume}{81}}, \bibinfo{pages}{31} (\bibinfo{year}{2002}).

\bibitem[{\citenamefont{Cleland et~al.}(2001)\citenamefont{Cleland, Schmidt,
  and Yung}}]{Cleland:2001}
\bibinfo{author}{\bibfnamefont{A.~N.} \bibnamefont{Cleland}},
  \bibinfo{author}{\bibfnamefont{D.~R.} \bibnamefont{Schmidt}},
  \bibnamefont{and} \bibinfo{author}{\bibfnamefont{C.~S.} \bibnamefont{Yung}},
  \bibinfo{journal}{Phys. Rev. B} \textbf{\bibinfo{volume}{64}},
  \bibinfo{pages}{172301} (\bibinfo{year}{2001}).

\bibitem[{Not({\natexlab{a}})}]{Note:Schmidt:2003}
\bibinfo{note}{D. R. Schmidt and C. S. Yung and A. N. Cleland, submitted to
  Appl. Phys. Lett.}

\bibitem[{\citenamefont{Little}(1959)}]{Little:1959}
\bibinfo{author}{\bibfnamefont{W.~A.} \bibnamefont{Little}},
  \bibinfo{journal}{Can. J. Phys.} \textbf{\bibinfo{volume}{37}},
  \bibinfo{pages}{334} (\bibinfo{year}{1959}).

\bibitem[{\citenamefont{Gantmakher}(1974)}]{Gantmakher:1974}
\bibinfo{author}{\bibfnamefont{V.~F.} \bibnamefont{Gantmakher}},
  \bibinfo{journal}{Rep. Prog. Phys.} \textbf{\bibinfo{volume}{37}},
  \bibinfo{pages}{317} (\bibinfo{year}{1974}).

\bibitem[{\citenamefont{Roukes et~al.}(1985)\citenamefont{Roukes, Freeman,
  Germain, Richardson, and Ketchen}}]{Roukes:1985}
\bibinfo{author}{\bibfnamefont{M.~L.} \bibnamefont{Roukes}},
  \bibinfo{author}{\bibfnamefont{M.~R.} \bibnamefont{Freeman}},
  \bibinfo{author}{\bibfnamefont{R.~S.} \bibnamefont{Germain}},
  \bibinfo{author}{\bibfnamefont{R.~C.} \bibnamefont{Richardson}},
  \bibnamefont{and} \bibinfo{author}{\bibfnamefont{M.~B.}
  \bibnamefont{Ketchen}}, \bibinfo{journal}{Phys. Rev. Lett.}
  \textbf{\bibinfo{volume}{55}}, \bibinfo{pages}{422} (\bibinfo{year}{1985}).

\bibitem[{\citenamefont{Wellstood et~al.}(1994)\citenamefont{Wellstood, Urbina,
  and Clarke}}]{Wellstood:1994}
\bibinfo{author}{\bibfnamefont{F.~C.} \bibnamefont{Wellstood}},
  \bibinfo{author}{\bibfnamefont{C.}~\bibnamefont{Urbina}}, \bibnamefont{and}
  \bibinfo{author}{\bibfnamefont{J.}~\bibnamefont{Clarke}},
  \bibinfo{journal}{Phys. Rev. B} \textbf{\bibinfo{volume}{49}},
  \bibinfo{pages}{5942} (\bibinfo{year}{1994}).

\bibitem[{\citenamefont{Nahum and Martinis}(1993)}]{Nahum:1993}
\bibinfo{author}{\bibfnamefont{M.}~\bibnamefont{Nahum}} \bibnamefont{and}
  \bibinfo{author}{\bibfnamefont{J.~M.} \bibnamefont{Martinis}},
  \bibinfo{journal}{Appl. Phys. Lett.} \textbf{\bibinfo{volume}{63}},
  \bibinfo{pages}{3075} (\bibinfo{year}{1993}).

\bibitem[{\citenamefont{Pothier et~al.}(1997)\citenamefont{Pothier, Gu\'{e}ron,
  Birge, Esteve, and Devoret}}]{Pothier:1997}
\bibinfo{author}{\bibfnamefont{H.}~\bibnamefont{Pothier}},
  \bibinfo{author}{\bibfnamefont{S.}~\bibnamefont{Gu\'{e}ron}},
  \bibinfo{author}{\bibfnamefont{N.~O.} \bibnamefont{Birge}},
  \bibinfo{author}{\bibfnamefont{D.}~\bibnamefont{Esteve}}, \bibnamefont{and}
  \bibinfo{author}{\bibfnamefont{M.~H.} \bibnamefont{Devoret}},
  \bibinfo{journal}{Phys. Rev. Lett.} \textbf{\bibinfo{volume}{79}},
  \bibinfo{pages}{3490} (\bibinfo{year}{1997}).

\bibitem[{Not({\natexlab{b}})}]{Note:Bandwidth}
\bibinfo{note}{The intrinsic electrical bandwidth of a tunnel junction is set
  by the product of the tunnel resistance $R_0$ and the junction capacitance
  $C_J$, $f_{\mathrm{3~dB}} = 1/2 \pi R_0 C_J$. For a fixed tunnel barrier
  thickness, this product is independent of the junction area $A$. With typical
  values of $R_0 A \sim 10^3~\Omega-\mu$m$^2$ and $C_J/A \sim 10^{-13}$
  F$/\mu$m$^2$, this corresponds to $f_{\mathrm{3~dB}} \sim 2$ GHz.}

\bibitem[{\citenamefont{Schoelkopf et~al.}(1998)\citenamefont{Schoelkopf,
  Wahlgren, Kozhevniv, Delsing, and Prober}}]{Schoelkopf:1998}
\bibinfo{author}{\bibfnamefont{R.~J.} \bibnamefont{Schoelkopf}},
  \bibinfo{author}{\bibfnamefont{P.}~\bibnamefont{Wahlgren}},
  \bibinfo{author}{\bibfnamefont{A.~A.} \bibnamefont{Kozhevniv}},
  \bibinfo{author}{\bibfnamefont{P.}~\bibnamefont{Delsing}}, \bibnamefont{and}
  \bibinfo{author}{\bibfnamefont{D.~E.} \bibnamefont{Prober}},
  \bibinfo{journal}{Science} \textbf{\bibinfo{volume}{280}},
  \bibinfo{pages}{1238} (\bibinfo{year}{1998}).

\bibitem[{not()}]{note:AlG}
\bibinfo{note}{The thermal conductivity of superconducting Al is exponentially
  supressed below $T_C$.}

\bibitem[{\citenamefont{Fulton and Dolan}(1987)}]{Fulton:1987}
\bibinfo{author}{\bibfnamefont{T.~A.} \bibnamefont{Fulton}} \bibnamefont{and}
  \bibinfo{author}{\bibfnamefont{G.~J.} \bibnamefont{Dolan}},
  \bibinfo{journal}{Phys. Rev. Lett.} \textbf{\bibinfo{volume}{59}},
  \bibinfo{pages}{109} (\bibinfo{year}{1987}).

\bibitem[{Not()}]{Note:Power}
\bibinfo{note}{Typically -100 dB (100 fW), the rf-SET of Ref
  \cite{Schoelkopf:1998} can tolerate much higher power than the rf-SIN}.

\bibitem[{not()}]{note:ROptimal}
\bibinfo{note}{The calculated value is at $R_{0} = L/C Z_{0} \sim 11$
  k$\Omega$}.

\bibitem[{Not({\natexlab{a}})}]{Note:SmallSignal}
\bibinfo{note}{To recover the small signal response for $n =$ 5, rewrite Eq.
  (\ref{eq:Tdot}) in terms of the reduced temperature $\epsilon \equiv
  (T_e-T_p)/T_p = \theta -1$ and retain terms $O(\epsilon)$, $\dot{\epsilon} =
  5 (\Sigma/\gamma)T_p^{n-2}\epsilon$.}

\bibitem[{Not({\natexlab{b}})}]{Note:WhichN}
\bibinfo{note}{For simplicity, we used $n=$ 5 for the fit.}

\bibitem[{\citenamefont{Deo et~al.}(2000)\citenamefont{Deo, Pekola, and
  Manninen}}]{Deo:2000}
\bibinfo{author}{\bibfnamefont{P.~S.} \bibnamefont{Deo}},
  \bibinfo{author}{\bibfnamefont{J.~P.} \bibnamefont{Pekola}},
  \bibnamefont{and} \bibinfo{author}{\bibfnamefont{M.}~\bibnamefont{Manninen}},
  \bibinfo{journal}{Europhys. Lett.} \textbf{\bibinfo{volume}{50}},
  \bibinfo{pages}{649} (\bibinfo{year}{2000}).

\bibitem[{\citenamefont{Nieuwenhuys}(1975)}]{Nieuwenhuys:1975}
\bibinfo{author}{\bibfnamefont{G.~J.} \bibnamefont{Nieuwenhuys}},
  \bibinfo{journal}{Adv. Phys.} \textbf{\bibinfo{volume}{24}},
  \bibinfo{pages}{515} (\bibinfo{year}{1975}).

\bibitem[{\citenamefont{Peters et~al.}(1984)\citenamefont{Peters, Buchal,
  Kubota, Mueller, and Pobell}}]{Peters:1984}
\bibinfo{author}{\bibfnamefont{R.~P.} \bibnamefont{Peters}},
  \bibinfo{author}{\bibfnamefont{C.}~\bibnamefont{Buchal}},
  \bibinfo{author}{\bibfnamefont{M.}~\bibnamefont{Kubota}},
  \bibinfo{author}{\bibfnamefont{R.~M.} \bibnamefont{Mueller}},
  \bibnamefont{and} \bibinfo{author}{\bibfnamefont{F.}~\bibnamefont{Pobell}},
  \bibinfo{journal}{Rev. Rev. Lett.} \textbf{\bibinfo{volume}{53}},
  \bibinfo{pages}{1108} (\bibinfo{year}{1984}).

\end{thebibliography}

\end{document}